\documentstyle[eqsecnum,aps,floats]{revtex}
\begin{document}
\tighten
\preprint{SU-GP-98/3-1,gr-qc/9803011}

\title{Summary of Session D1(i), Quantum General Relativity}

\author{Donald Marolf}
\address{Physics Department}
\address{Syracuse University, Syracuse, NY 13244}

\date {February, 1998}

\maketitle

\section{Introduction}

Looking back over the list of speakers from the session on {\it Quantum
General
Relativity}, the most striking feature is the huge breadth covered by the 
the two afternoons (and 17 speakers!) of the session.  Speakers from
ten different countries spoke on topics ranging from measures to mapping 
class groups, from anomalies to asymptotic structure, and from strings 
to solitons.  Clearly, there is not sufficient space here to discuss
all of these works in detail, but I have tried to include at
least a few comments on each talk, together with appropriate references.
Many thanks to the speakers for sending me brief summaries of their 
presentations from which to work.

I will attempt to group the talks by subject, though close
relations between the talks will not exist in general.  The one clear
exception is in the set of talks on loop gravity
(by Thiemann, Lewandowski, Loll, and
Pullin), which I will discuss at the end of this summary.

\section{Summaries}

The session opened with ``The measure in Simplicial Gravity'' by Ruth Williams,
giving a
solid  review of what is known about that issue.
A good introduction to this subject can be found in Williams' previous 
review \cite{RW} with H.W. Hamber.
  Whether a measure can be chosen so that simplicial gravity
has
the proper continuum limit is a question under active investigation, but a
necessary
criterion for this to happen is for the lattice measure to agree with its
continuum 
counterpart in the limit of weak fields and low momentum.  Williams uses this
to argue that
one should carefully study discretizations of the continuum measure.  This is
in contrast
to another strategy in which a discrete form of the supermetric is sought
first, and then
the determinant of this discretized metric is evaluated.  Other important
concerns, such
as those based on locality and on factorization and composition laws, were also
discussed.
Williams' was the only presentation to discuss discretized approaches to
quantum gravity.

The work ``Models of Coherent State Evolution in Quantum Gravity''
presented by Michael Ryan was a continuation of his interest in the question
of
to what extent quantum
minisuperspace models could actually approximate a full theory of  quantum
gravity \cite{RK}.
The work presented in Pune 
was part of an on-going program of research, but some
earlier results and background may be found in \cite{MR}.  The idea was to use
coherent
states to probe related issues in $\lambda \phi^4$ field theories.  
In $\lambda \phi^4$ theory on a compact space, the homogeneous sector
can play the part of a minisuperspace model.
The issue of course is,
if one begins with a coherent state of the full theory which well approximates
a state in the minisuperspace model, whether the evolution of this state will 
then track the 
evolution of the minisuperspace state  or whether, due to nonlinearities and
dispersion, 
the evolutions will rapidly diverge. 
He is also currently pursuing such issues in
other field theories, such as the Sine-Gordon model, 
which can be directly related
to 1+1 gravity.

The connection between sine-Gordon
theory and 1+1 gravity was in fact the subject
of another talk, ``Black Holes and Solitons''
by Gabor Kunstatter.  He discussed work \cite{GK} showing the
connection between black holes in (Lorentzian signature) Jackiw-Teitelboim
gravity
and Euclidean sine-Gordon solitons.  It was 
previously known that the metrics
that solve this theory also solve the sine-Gordon equation in an appropriate
sense.
Kunstatter and his collaborator Gegenberg have now shown that the dilaton,
which
generates Killing vectors in this gravity theory, also generates symmetries of
the
sine-Gordon soliton.  While the gauge structure of the two theories is vastly
different
on the surface, Kunstatter expressed the hope that this connection could lead
to a 
better understanding of black hole entropy, perhaps following the ideas of
Carlip
\cite{SC}.

Kunstatter's was not the only talk to touch on black hole entropy.  Hans
Kastrup
also discussed his ideas \cite{HK} for studying the entropy of black holes
by first canonically quantizing the (3+1) Schwarzschild metric in a certain
way
and then relating the resulting states to the Ising  droplet nucleation model
in 2 dimensions.  His talk ``Quantum Statistics of Schwarzschild Black Holes
and Ising Droplet Nucleation'' also showed that this connection can be extended
to higher dimensional
black holes and higher dimensional droplet models.

The canonical quantization of black holes in particular and of spherically
symmetric gravity more generally was discussed in some depth by
John Friedman.  Friedman's talk ``Topological Geons in the context of
spherically symmetric
minisuperspace'' was submitted with two coauthors, Jorma Louko and 
Steven Winters-Hilt, and the work of all three was discussed.  They were
interested
both in vacuum wormholes of various topologies and in such spacetimes with an 
additional degree of freedom corresponding to a collapsing shell of massive 
dust \cite{JF}.
Using the quantizations they found to be natural (associated, for example,
with
the proper time along the shell), they found a discrete energy spectrum for
the
black hole.   

Another talk concerning black holes, ``Non-commutative Black Hole Algebra
and String Theory from Gravity,'' was given by Sebastian de Haro Olle. 
Using the framework of 't Hooft, he \cite{SO} generalized an action which
describes gravitational scattering between point particles and found that it
has just the form of a string theory action, complete with an antisymmetric
tensor term and a dilaton.   At the quantum level, he found that this
leads to four noncommuting coordinates.

The final talk related to black holes black holes was given by Sukanta Bose.  
While the black holes themselves were not his central topic, his talk, ``On
different approaches to quantizing two-dimensional dilaton-gravity models" 
concentrated on models that are famous for containing
1+1 dimensional black holes.  The point of this work was to compare
the perturbative
path integral quantization of such models with an exact canonical
quantization.
He found that the anomaly in the canonical commutator algebra was related
to the Polyakov-Liouville term of the path integral method, and that the
choice of canonical anomaly potential is related to the choice of
local covariant counterterms
in the one-loop action and to the quantum state of the matter fields.
The talk is based on unpublished work, but see \cite{SB} for 
related work by Bose.

A second talk on 1+1 dimensional theories was ``Quantum Fields at
any time by Madhavan Varadarajan, for which Charles Torre was a coauthor.
The purpose of this delightful talk
was to address issues of unitarity when studying
quantum fields propagating on curved spacetime backgrounds.  In fact, 
they studied only flat backgrounds, but considered arbitrary slicings.
They asked the question of whether the 
fields (and thus the Schr\"odinger picture Hilbert space) associated
with some given hypersurface are unitarily related to those of an
arbitrary hypersurface.  They find that such evolution is unitary
on the manifold ${\bf R} \times S^1$ \cite{MV}, but that it is generically
not so on ${\bf R} \times {\bf R}$ or on higher dimensional manifolds. 
The phenomenon is analogous to the lack of unitarity that arises
in curved spacetime
quantum field theory due to an infinite creation of particles.

A third lower dimensional talk was 
Jeanette Nelson's ``Constants of motion and
the conformal anti-De Sitter algebra in (2+1)-Dimensional Gravity.''  
Nelson reported on work \cite{JN} with
her collaborator, Vincent Moncrief, in which constants of motion were
calculated
for the case where the spacetime topology is ${\bf R} \times T^2$.
The algebra of these observables 
was related to the conformal group and 
the action of large diffeomorphisms on this algebra was discussed.

Large Diffeomorphisms were also
addressed by Dominico Giulini in his talk ``Mapping Class Groups 
and their reduction in Quantum Gravity.''  He was interested in the
structure of the mapping class groups of
three dimensional manifolds and of how to 
reduce a classical or quantum observable algebra with respect to
such diffeomorphisms.
He reported a number of results concerning 
manifolds that can be represented as connected sums of prime manifolds.  
When the connected sum contains no handles ($S^1 \times S^2$), the
mapping class group has
the structure of an iterated semi-direct product.
He also showed that, for a large class of
(and perhaps all) three-manifolds, this group is residually finite.   Some
of Giulini's
past work on this subject can be found in \cite{DG}.
 
On a completely different note, Carlos Kozameh presented work
on ``The phase space of Radiative Spacetimes'' which was done together
with Malcolm Ludwigsen.  Some of this work has already appeared
\cite{CK}.   By studying the covariant symplectic structure of general
relativity, 
they were able to show that the phase space of GR is in fact foliated into
an infinite number of leaves, each labeled by the value of the mass 
aspect ($\psi_2$) at spatial infinity.  In a quantum theory, these leaves
would correspond to superselection sectors.  The ensuing discussion 
suggested that, as usual, these superselection rules may be related to
gauge transformations that are not generated by
constraints in the corresponding 
canonical theory.

The remaining non-loop talk, ``Global Anomalies in Canonical Gravity,'' was
given by Sumati Surya.
She and her collaborator, Sachindeo Vaidya, were
interested in using topological arguments
to study the possibility of certain gravitational anomalies that might
arise in the presence of chiral fermions.
The point here is
that known anomalies in theories with chiral fermions
can be related to the existence of non-contractable curves in what, in
the GR context, would be
called the space of semi-classical theories; that is, in the space
of theories defined by
quantizing matter fields on a given classical background.  The anomalies
arise when the adiabatic fermion vacuum changes by a nontrivial phase under
transport around such a loop.
For technical reasons, they
introduced an SU(2) gauge field and found that, when coupled
to gravity on a manifold $\Sigma \times {\bf R}$, all such phases
are trivial when 
$\Sigma$ is a Lens space \cite{SS}.  As a result, such
a theory should be free of this kind of anomaly.

The remaining five talks concerned the loop approach to quantum gravity.  Since
these were in fact
closely related, I will discuss them together.  First, however, it is
appropriate to simply list
the speakers and their titles:  Thomas Thiemann -- ``Towards Solving the
Quantum
Einstein Equations: A Status report,''
Roberto de Pietri -- ``The Matrix Elements of the Constraint Operators
in Loop Quantum Gravity,'' Jerzy Lewandowski -- ``Loop constraints: A habitat,
their algebra, and
solutions,'' Renate Loll -- ``Lattice Methods as a Tool for Understanding
Quantum Effects,''
and  Jorge Pullin -- ``The conflict between diffeomorphism invariance and the
field theoretic
nature of quantum gravity.''

A common theme of many of these talks was the constraints
and dynamics of quantum (loop)
gravity.  Thiemann gave an overview of his past work \cite{TT} on
the subject and of his current thoughts for constructing improved constraints.
De Pietri
discussed some details of the matrix elements of constraints that have
been proposed by Thiemann \cite{QSD} and a related, but perhaps alternative
formulation of loop gravity through path integrals which has been described
by Reisenberger and Rovelli \cite{RR}.  For De Pietri's past work on the
subject, see
\cite{RdP}.

Lewandowski
discussed some difficulties that have recently been found \cite{JL} with past
proposals
for the constraints.  In particular, the algebra of these constraints appears
to be a rather
truncated version of the hypersurface deformation algebra.   He also discussed
a general incompatibility between having well-defined Hamiltonian constraints
and
having as much Hilbert space structure as one might like.  The point here
is that if the Hamiltonian constraints were well-defined operators in some
Hilbert
space and if their domains contained states $|\psi \rangle$ 
which were invariant under spatial diffeomorphisms, 
then the inner product $\langle  \psi |H(N) H(M)|\psi \rangle$ would provide a
diffeomorphism
invariant bilinear form on the fields $N$ and $M$.  However, the first paper 
of \cite{JL}
shows that, unless $N$ and $M$ are
densities of weight $1/2$, the only such
bilinear form is the trivial one (zero).

The talks of Thiemann and Pullin also referred to
the constraint algebra difficulties
found in \cite{JL}, and both speakers expressed the belief that the
problem could
be avoided by some modification of the loop approach.
Thiemann favored changing the definition of spin network states to allow an
infinite number of vertices, while Pullin suggested that more attention should
be paid to results associated with the Chern-Simons state and certain knot
invariants (the Vassiliev invariants) that are associated with Chern-Simons
theory.
He and Rodolfo Gambini have introduced \cite{JP}
a well-defined
loop derivative on the space of such invariants and have
generalized the notion of Vassiliev
invariants to spin-networks.  The idea is that this may allow the development
of a new class of proposals for the Hamiltonian constraint. 

The talk by Loll had a somewhat different flavor.  While inspired
by loop techniques, she is interested in the idea that the discreteness
present in the loop approach may not be fundamental in an of itself, 
but may form a useful approximation to a continuum limit.  In this
context, for example, the discreteness of areas and volumes in
loop quantum gravity may also not be fundamental.  She reported
the results of a recent calculation \cite{RL} suggesting that the 
quantum algebra of discrete diffeomorphisms in such an approach
closes in the limit of vanishing lattice spacing.

\end{document}